\documentclass[aps,prl,twocolumn,superscriptaddress,floatfix]{revtex4-2}
\usepackage{multirow}
\usepackage[latin9]{inputenc}
\usepackage{times}
\usepackage{amssymb}
\usepackage{amsmath}
\usepackage{graphicx}
\usepackage{graphics}

\usepackage{upgreek}

\usepackage{color}
\usepackage{bm}

\makeatletter

\begin{document}

\title{
Observation of period-doubling Bloch oscillations}

\author{Naveed Khan}
\thanks{These authors contributed equally to this work}
\affiliation{School of Physics and Astronomy, Shanghai Jiao Tong University, Shanghai 200240, China}

\author{Peng Wang}
\thanks{These authors contributed equally to this work}
\affiliation{School of Physics and Astronomy, Shanghai Jiao Tong University, Shanghai 200240, China}

\author{Qidong Fu}
\email[]{Corresponding author: jellyfqd@sjtu.edu.cn}
\affiliation{School of Physics and Astronomy, Shanghai Jiao Tong University, Shanghai 200240, China}

\author{Ce Shang}
\affiliation{Physical Science and Engineering Division (PSE), King Abdullah University of Science and Technology (KAUST),  Thuwal 23955-6900, Saudi Arabia}
 
\author{Fangwei Ye}
\email[]{Corresponding author: fangweiye@sjtu.edu.cn}
\affiliation{School of Physics and Astronomy, Shanghai Jiao Tong University, Shanghai 200240, China}
\date{\today}

\begin{abstract}

Bloch oscillations refer to the periodic oscillation of a wavepacket in a lattice under a constant force. Typically, the oscillation has a fundamental period that corresponds to the wavepacket traversing the first Brillouin zone once. 
Here we demonstrate, both theoretically and experimentally, the 
optical Bloch oscillations where the wavepacket must traverse the first Brillouin zone twice to complete a full cycle, resulting in a period of oscillation that is two times longer than that of usual Bloch oscillations. The unusual Bloch oscillations arise due to the band crossing of valley-Hall topological edge states at the Brillouin boundary for zigzag domain walls between two staggered honeycomb lattices with inverted on-site energy detuning, which are protected by the glide-reflection symmetry of the underlying structures. Our work sheds light on the direct detection of band crossings resulting from intrinsic symmetries that extend beyond the fundamental translational symmetry in topological systems.

\end{abstract}

\maketitle

Bloch oscillations (BOs) are periodic oscillations exhibited by a wavepacket in a periodic lattice under a constant force. The oscillations have a period that takes the wavepacket to transverse the first Brillouin zone once. As such, BOs offer a valuable tool for revealing crucial information about the band structure and transport properties of periodic systems. Initially predicted for electrons in crystalline lattices~\cite{bloch1928quantum,zener1934theory} and shortly after the observation of Wannier-Stark ladders~\cite{mendez1988stark,voisin1988observation}, BOs were observed first in semiconductor superlattices~\cite{feldmann1992optical,waschke1993coherent}. Since then, they have been observed in various systems including ultracold atoms~\cite{dahan1996bloch}, Bose-Einstein condensates(BECs)~\cite{morsch2001bloch, anderson1998macroscopic}, waveguide arrays~\cite{peschel1998optical,pertsch1999optical,stutzer2013hybrid,joushaghani2009quasi, longhi2009prl}, optically induced lattices~\cite{trompeter2006bloch,sun2018observation}, plasmonic waveguides~\cite{block2014bloch}, and systems with parity-time symmetry~\cite{wimmer2015observation}. BOs have also been observed in strongly correlated quantum systems, but with a doubled frequency~\cite{prb_nlbo,science_nlbo,nc_nlbo}.

 BOs occur in topological materials too~\cite{lu2014topological}. These materials have a nontrivial topological structure in their Bloch bands, defined by certain invariants, and their accurate measurement is crucial to characterize the topological materials. In this regard, BOs have proven to be a powerful tool for measuring these invariants, such as Chern numbers or Zak/Berry phases~\cite{atala2013direct,aidelsburger2015measuring,jotzu2014experimental,xiao2010berry}. Interestingly,  in truncated topological structures~\cite{epl2014,li2019bloch,plotnik2015topological} or  for spin-orbit coupled atoms loaded in a one-dimensional (1D) Zeeman lattice~\cite{kartashov2016bloch}, it is predicted that completing one BO cycle  requires scanning the Brillouin zone twice, resulting in a period doubling that of usual BOs ~\cite{epl2014,li2019bloch,plotnik2015topological}.  
However, realizing these predictions is challenging due to the need for strong time-reversal-symmetry breaking, and the practical difficulties in implementing them in 1D system ~\cite{kartashov2016bloch} or transforming edge states with delocalized bulk modes. Additionally, by bypassing constraints in real space, indirectly achieving period-doubling BOs for BECs on a synthetic topological Hall cylinder has been mentioned~\cite{virtualBO}. However, this was done in a virtual, synthetic space, hence a direct observation of period-doubling BOs has not been achieved yet.

Here we investigate both theoretically and experimentally, the optical BOs with a doubling period without breaking time-reversal symmetry. The BOs occur on valley-Hall edge states (VHESs) at a domain wall between two inverted staggered photonic honeycomb lattices. The system preserves the time-reversal symmetry, enabling wavepackets to move in either direction within the same domain wall. 
We consider two types of domain walls: zigzag and bearded. While both types support a pair of localized states due to the valley-Hall effect, the zigzag domain wall has energy band crossings for VHESs, resulting in the need to scan the Brillouin zone twice for the BOs of zigzag VHESs. This is in contrast to the bearded domain wall or other usual BOs, where one cycle is achieved by traversing the Brillouin zone once. Unlike previous proposals on period doubling of BOs~\cite{epl2014,li2019bloch,plotnik2015topological,kartashov2016bloch}, our study considers BOs in a 2D system, where the back-and-forth oscillation occurs within the same domain wall without involving bulk mode transformations. This facilitates the experimental measurement of the wavepacket drift in real space.  By tracking the ``center of mass" (COM) of the VHESs, we directly observed, to the best of our knowledge, the first period-doubling BOs in photonic systems or any other physical systems.

We analyze the light propagation along the $z$-axis in a photorefractive medium, specifically  strontium barium niobate (SBN: 61)  that is  described by the following dimensionless equation for the light amplitude $\Psi$:
\begin{equation}
i\frac{\partial\Psi}{\partial z}=-\frac{1}{2}(\frac{\partial^2}{\partial x ^2}+\frac{\partial^2}{\partial y ^2})\Psi-
\frac{E_{0}}{1+I_{L}(x,y)+I_{G}(y)}\Psi,
\label{eq:1}
\end{equation}
where the applied dimensionless d.c. field is set to $E_0=-30$, corresponding to an electric field $E$ of 2.4$\times$10$^5$ Vm$^{-1}$.  $I_L (x, y)$ represents a 2D lattice wave field that consists of a domain wall and two staggered honeycomb lattices with inverted on-site potentials surrounding it. Two types of domain walls are considered: zigzag [Fig.~1(a)] and bearded [Fig.~1(c)] (see~\cite{supplemental} on how these lattices are constructed in modelling and in experiment). To induce BOs, a potential gradient is created along the domain walls by illuminating the sample laterally with a gradient white light, represented by $I_G =I_0 $[1+tanh($y/\eta$)], where $I_0$ is the intensity of the white light, and $\eta$ characterizes the width of its intensity gradient that controls the rate $\beta$ of a linear increase in the refractive index in the $y$ axis~\cite{supplemental}. Note that the form in which the lattice intensity $I_L(x,y)$ and gradient light intensity $I_G(y)$ enter Eq. (\ref{eq:1}) is determined by the mechanism of the photorefractive response~\cite{prc1,prc2}.

To investigate the impact of the Floquet-Bloch spectrum on the dynamics of BOs, we first set the gradient $I_G$ to zero and search for Bloch eigenmodes in the form of $\Psi(x,y,z) = \psi(x,y)e^{iEz}=u(x,y)e^{i k_y y+iEz}$, where $E(k_y)$ represents the energy bands, $k_y \in [-K/2, K/2]$ is the Bloch momentum along the $y$ axis, $K=2\pi/a_y$ is the width of Brillouin zone, $a_y$ is the lattice's y-periodicity, and $u(x,y)\,=\,u(x,y+a_y)$ is the periodic part of the Bloch function $\psi$.

\begin{figure}[t]
\centering
\includegraphics[width=0.99\columnwidth]{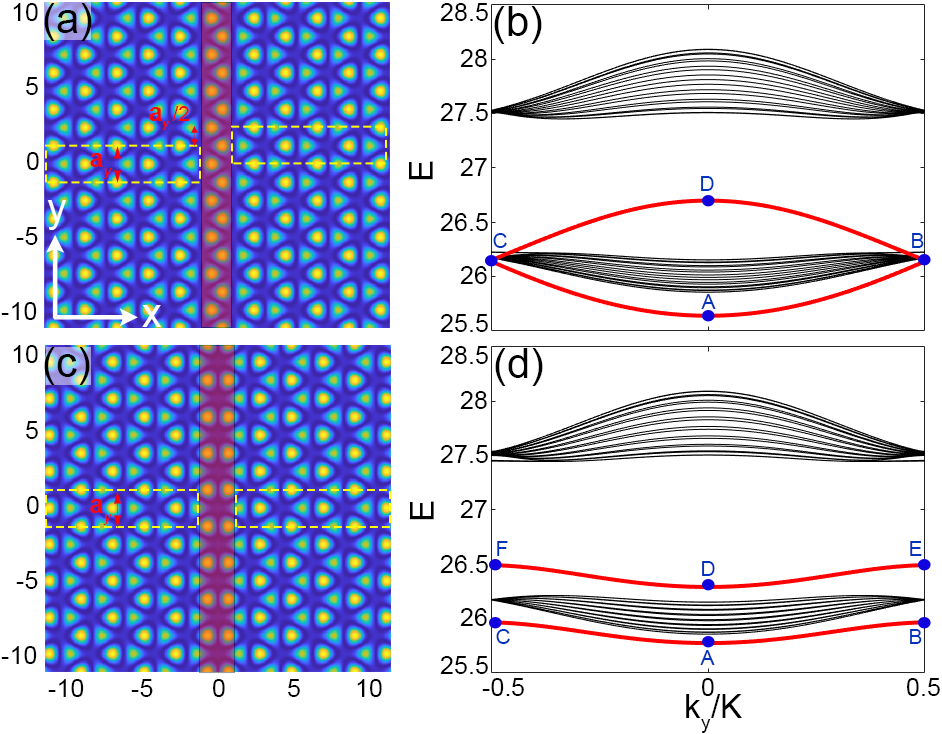}
\caption { (a, c) Domain walls of zigzag-type (a) and bearded-type (c), between two inverted, staggered honeycomb lattices. The red-shaded regions represent the domain walls, and the rectangular dashed yellow boundaries on the left and right of the domain walls are used to illustrate the glide-reflection symmetry in (a) and mirror symmetry in (c). (b, d) Band structures of the lattices corresponding to (a, c). Red curves represent the VHESs, while the black curves represent bulk states. Blue dots on the red curves represent VHESs at the center ($k_y=0$) or boundaries ($k_y=\pm K/2$) of the Brillouin zone respectively. Note that VHESs (red curves) are localized in the $x$
direction but the bulk states (black curves) are delocalized in the
$x$ direction, and in what follows we are going to excite only
the localized states.
}
\label{fig1}
\end{figure}

The spectrum $E(k_y)$ is displayed in Fig.~1 (b) and (d). As expected, due to the breaking of the spatial inversion symmetry in the staggered honeycomb lattices, a gap opens and a pair of VHESs emerge (indicated by the red curves) ~\cite{noh2018observation}. However, the VHESs exhibit essentially distinct spectra rooted in the lattice symmetries.
The lattice with the bearded domain wall has mirror symmetry $\mathcal{M}_x (x \rightarrow -x)$ along the $x$-axis [Fig.\ref{fig1}(c)]. Applying $\mathcal{M}_x$ twice returns a field component to its original, namely, $\mathcal{M}_x^2 \psi(x,y)= \psi(x,y)$. Correspondingly, $\mathcal{M}_x \psi(x,y)= \pm \psi(x,y)$, resulting in the gapped VHESs, as shown in Fig.\ref{fig1}(d). On the other hand,  the lattice with the zigzag domain wall lacks a straightforward mirror symmetry $\mathcal{M}_x$ but has a glide-reflection symmetry denoted as $G=\mathcal{M}_x T_y(a_y/2)$, namely, the lattice is invariant under $\mathcal{M}_x$ followed by a translation along the $y$ direction by half a lattice constant. When $G$ is applied twice, it results in a full-lattice-constant translation, i.e.,  $G^2 \psi(x,y)= \psi(x,y+a_y)$. By Bloch's theorem, $G \psi(x,y)= e^{ik_ya_y/2}\psi(x,y)$. Considering $k_y=\pm K/2$, such that $G^2=-1$, one finds that the eigenvalues of $G$ are $\pm i$~\cite{PhysRevB.81.155115}, and consequently, bands must cross by pairs at $k=\pm K/2$, as confirmed in Fig.~1(b). This band crossing is protected by the glide-reflection symmetry and is insensitive to a limited continuous deformation of the lattice that preserves the symmetry~\cite{glidewave, note}. Note that, band crossings can also be achieved in tilted lattices by delicately balancing the coupling between sites and external forces~\cite{tilt0,tilt1, tilt2}. However, symmetry-protected band crossing facilitates experimental observations of the associated band dynamics, with BOs being a salient example, as will be shown below.

With a potential gradient $I_G$ present in Eq.~\ref{eq:1}, the VHES is subject to a constant force along $y$. If the force is small,  the evolution of the state is adiabatic, so that it remains in the same energy band and doesn't transition between bands. The Bloch momentum of the VHES increases linearly with $k_{y}(z) = k_{y}^0 + \beta z$, $k_y^0$ being the initial momentum.  
For VHESs in a bearded domain with initially zero momentum, they follows either the path $\textbf{\text{A$\rightarrow$B}$\rightarrow${C$\rightarrow$A}}$ or $\textbf{\text{D$\rightarrow$E}$\rightarrow${F$\rightarrow$D}}$ [Fig. 1(d)],  scanning the whole Brillouin zone once and returning to the initial state. Therefore, the BOs period in momentum space for the bearded state is $K$, corresponding to a distance $Z=K/\beta$ in real space. This is the typical BOs scenario.

The situation changes significantly when it comes to the zigzag VHESs due to band crossing that directly results in the period doubling of BOs.
Indeed, consider a VHES with zero momentum $k_y=0$ in the lower band at point $\textbf{A}$ in Fig.~1(b). Due to the constant force, it moves up to point $\text{\textbf{B}}$ at the Brillouin border where the lower band crosses with the upper band. Upon crossing, it reappears at the opposite border $\text{\textbf{C}}$, but now on the upper band. It does so because the phase distributions of the VHESs or the pseudospin, as defined in relation to it, precisely match between the upper and lower bands at the Brillouin zone border~\cite{supplemental}. Conversely, the phase or pseudospin of the VHESs at the two edges of the same band is completely opposite (orthogonal), preventing movement along the same band across the Brillouin zone. 

Therefore, after the band crossing, the VHES at point $\text{\textbf{C}}$ continues
moving up along the upper band until it reaches the band edge at point  $\text{\textbf{B}}$, where it undergoes another band crossing and moves back to the lower band, and ultimately return to its initial position $\text{\textbf{A}}$. 
Thus, unlike the bearded domain case, where only one band is involved throughout the BOs, the zigzag domain involves sequential traversal of two bands. This means that they must traverse the Brillouin zone twice,  or a total distance of $2K/\beta$ to complete a full BOs.

The aforementioned analysis was confirmed through direct simulations of propagation using Eq.~(1), with an excitation condition being a VHES that is horizontally localized and vertically modulated by a broad Gaussian envelope of width $w=8$ and centered at, say,
$y=y_0=-6$, expressed as $\Psi(x,y)|_{z=0}=\psi(x,y; k_y)\exp[-({y-y_0})^2/w^2]$. Specifically, we selected the Bloch wave with momentum $k_y=0$ corresponding to point $\text{\textbf{A}}$ in Fig.~1(b) and (d). Such selected VHESs initially move along the domain wall in the positive $y$ direction [Fig.~2(a), (b)]. After reaching their maximum displacement, they return to their initial positions, and a new cycle starts. To quantify the oscillation, the variation of COM of the wavepacket, defined by $y_c=\iint {\Psi}^{*} y {\Psi}dx d y$, was monitored [Fig.~2(c)]. 
Our results clearly show that while the bearded state completes one full BOs at $z=27$, the counterpart wavepacket in the zigzag domain only reaches its maximum displacement, i.e., completing only half of its BO cycle. Subsequently, it takes an additional distance of $27$ for the wavepacket to return to its initial position and complete one full oscillation cycle. By this distance, the wavepacket in the bearded domain has already completed two full oscillations. 
Note also, as the amplitude of BOs is proportional to the width of the involved band(s)(divided by the force $\beta$), the amplitude of two-band BOs in the zigzag domain wall is much larger (around 4 times larger in this case) than that of one-band BOs in the bearded one.
\begin{figure}
\centering
\includegraphics[width=1\columnwidth]{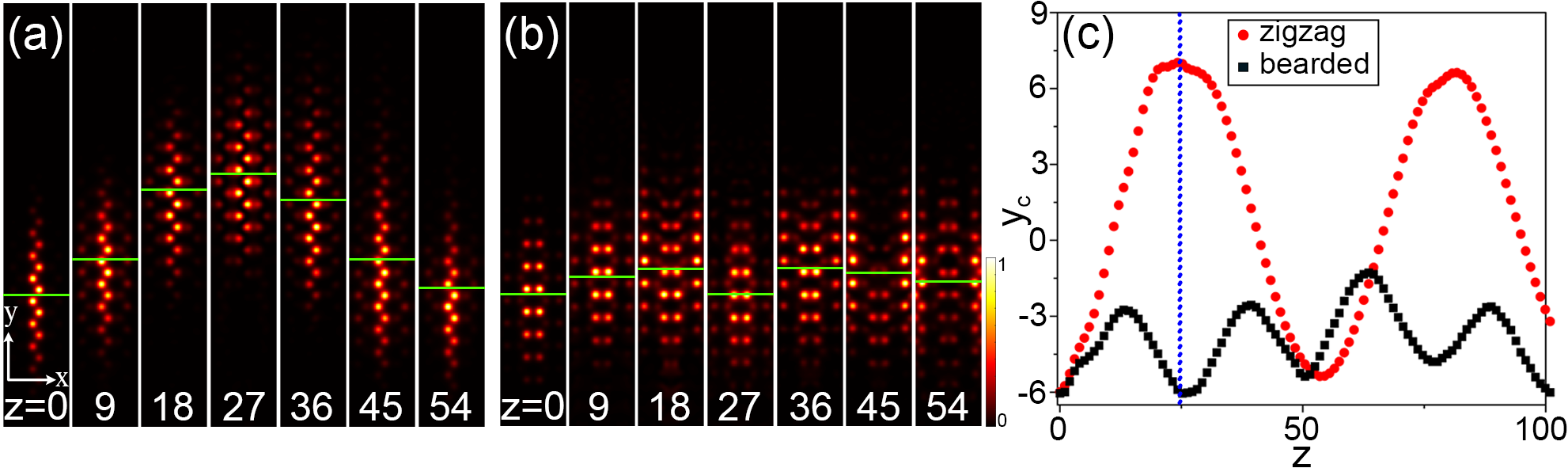}
\caption{Simulation results for the VHESs evolving at the zigzag (a), and (b) bearded domain walls, respectively. In both figures, the horizontal green lines indicate the location of COM ($y_c$) of the light beam, that are further shown in (c). 
The dashed vertical blue lines shown in (c) indicate the point where the light beam along the zigzag domain wall attains its maximum displacement, which is equivalent to half of its period of BOs.
$E_0=-30$, $\eta=35, I_0=0.25$, $w=8$.}
 \label{fig2}
 \end{figure}

\begin{figure}[t]
\centering
\includegraphics[width=1\columnwidth]{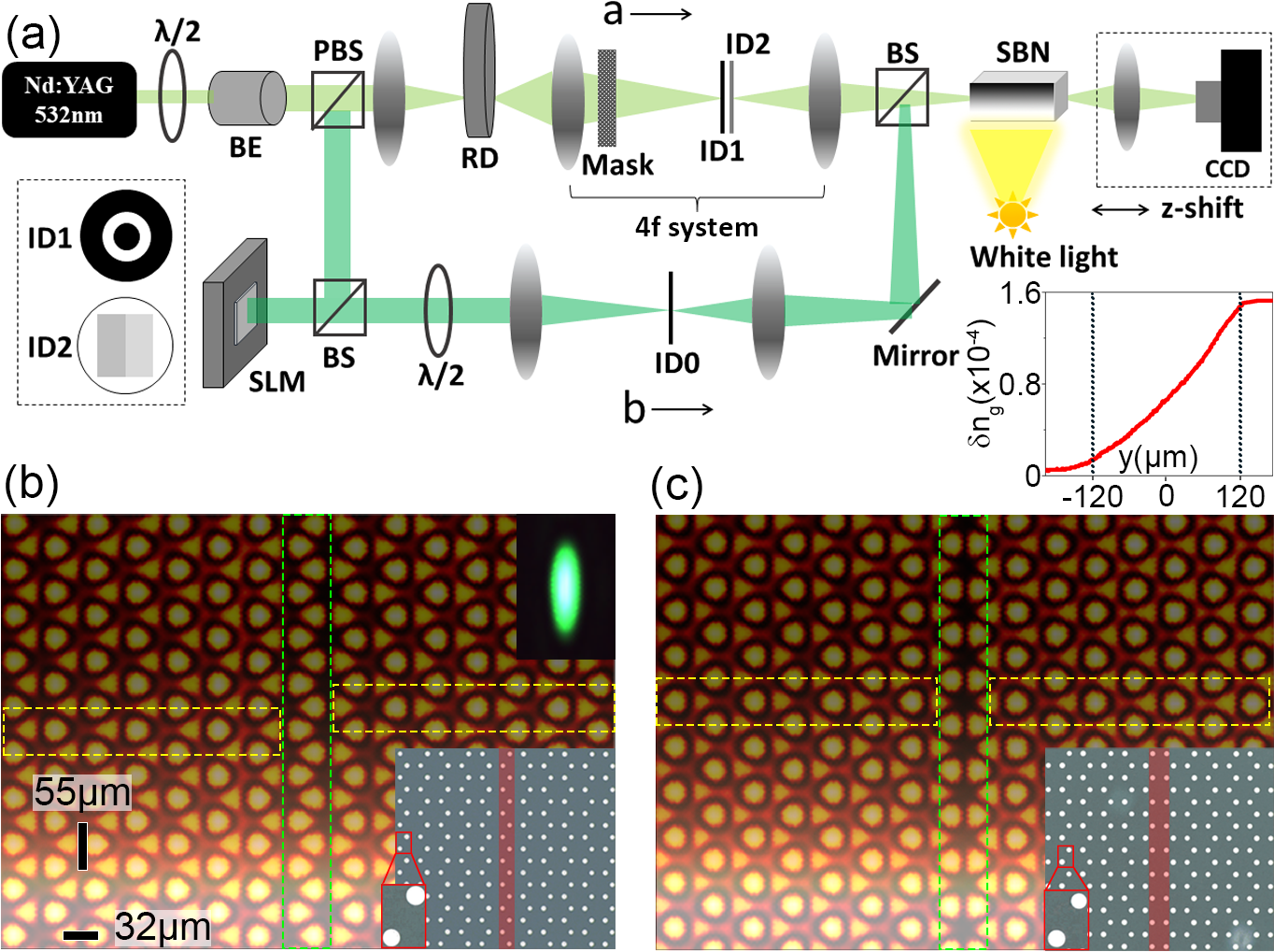}
\caption{(a) Schematic diagram of the experimental setup. $\lambda/2$: half-wave plate; PBS: polarization beam splitter; BS: beam splitter; BE: beam expander; RD: rotating diffuser; ID0: iris diaphragm; ID1: Fourier filter with six holes; ID2: phase mask; SLM: spatial light modulator. The panel at the bottom-right corner shows an example of a refractive index ramp $\delta n$ with a width $\eta=120~\mu\text{m}$.
(b, c) The lattice-forming light field measured at the output of the SBN, showing staggered honeycomb lattice profiles with zigzag (b) and bearded (c) domain walls. The domain walls are depicted within green rectangular-dashed boundaries. The brightness of the color in the lattices is proportional to the local light intensity. The insets at the bottom-right corners of (b) and (c) show the designed structures on the amplitude mask, while the inset at the top-right of (b) displays the probe light beam at the front facet of the crystal. }
\label{fig3}
\end{figure}

We experimentally investigated BOs with the setup in Fig.~\ref{fig3}(a). A Nd: YAG laser beam ($\lambda=532$\, nm) was split into two beams: beam $a$, with ordinary polarization, writing the required lattice structure into SBN, while beam $b$, with extraordinary polarization,  probing the lattice and studying the BOs. 
Note that the crystal's large electro-optic anisotropy was utilized in the polarization configuration, to ensure that the probe experiences strong modulation of the lattice, but the induced lattice remains undistorted during propagation~\cite{prc1,prc2}.
 The writing beam passed through a rotating diffuser, which made it partially spatially incoherent, and then went through an amplitude mask that has a pattern matching the structure of the desired writing beam, as shown in the bottom-right corner of Fig.~\ref{fig3} (b) and (c). The mask was then imaged onto the input facet of the crystal through a 4f optical system with an appropriate filter at the Fourier plane (see the experimental details in~\cite{supplemental}). This process generated an unconventional lattice-forming beam that remains nondiffracting over a distance of at least $2$ cm in the sample~\cite{martin2004discrete,chen2002spatial}.  
 Furthermore,  to introduce a ramp of refractive index across the sample, a white light diffracted by a sharp razor blade was utilized to laterally illuminate the SBN~\cite{trompeter2006bloch,sun2018observation,terhalle2011dynamic}. 

Figures \ref{fig3}(b) and 3(c) show the optically induced staggered honeycomb lattices with zigzag and bearded domains, overlaid with a refractive index gradient.
The total induced refractive index change is given by $\delta n=-\gamma/(1+I_L+I_G)$, where $\gamma$ = $ {n_e^3}{r_{33}}E/2 $ =  $ 3.56\times$10$^{-4}$ and $ {r_{33}} $ is the nonlinear electro-optic coefficient. The intensity of the lattice-writing beam is $\ I_L \approx$  $ 2.5 ~\text{mW}/\text{cm}^2$,  and the maximum intensity of the white light is $I_G \approx $ $ 4.2 ~\text{mW}/\text{cm}^2$. An example of the refractive index ramp $\delta n_g$ is given in Fig.~\ref{fig3} (a). As shown, the index change over one period $a_y=55~\mu$m due to the ramp is $\sim 10^{-5}$, one order of magnitude smaller than the variation of index of the periodic lattice $\sim 10 ^{-4}$. This implies that the influence of the gradient on the profiles of Bloch modes is negligible, and its main effect is to initiate the variation of Bloch momentum of the wave packet in the Brillouin zone.

\begin{figure}[htb]
\centering
\includegraphics[width=1\columnwidth]{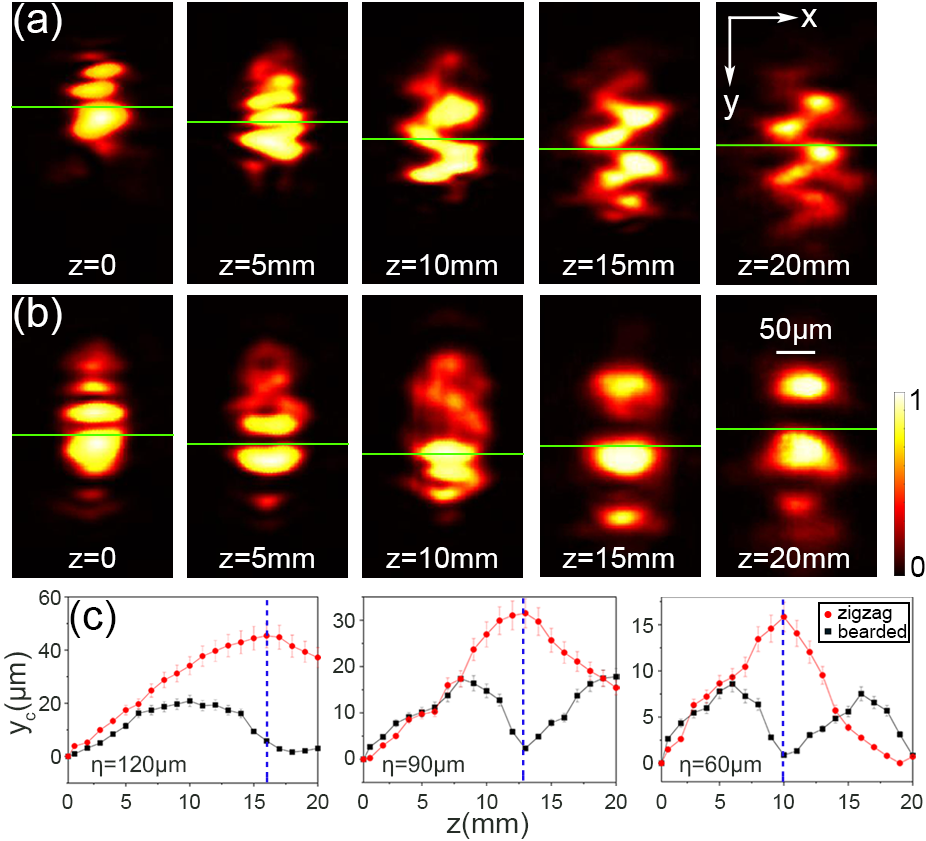}
\caption{Experimental observation of BOs at (a) zigzag and (b) bearded domain walls. The horizontal green lines indicate the position of COM of the light beam. The index gradient applied in the lattices increases in the downward direction, and for the shown particular case, the width of the index gradient is  $\eta=120 ~\mu\text{m}$. Note in the bearded state case (b), the close proximity of two horizontal potential maxima makes the light spots between them unresolvable [cf Fig. 2(b)]. (c) Displacement of the COM versus propagation distance $z$ under three different index gradients. 
}
\label{fig4}
\end{figure}

The lattices are excited by launching into both domain walls a Gaussian laser beam with an elliptical shape that corresponds to the numerical simulations. The beam has a width of $90~\mu \text{m}$ along the domain wall extension ($y$-axis), and $30~\mu \text{m}$ across the domain wall ($x$-axis). Its elliptical shape ensures that it is wide enough along the $y$ direction so that only one Bloch mode at $k_y=0$ is excited,
 yet narrow enough to fit into the width of the domain wall. Thanks to the topological protection of the valley Hall effect, both types of the VHESs are efficiently excited, as evidenced by their quick reshaping into the ``zigzag" and ``bearded" profile, and remain well-trapped in the domain wall while moving along it, without radiating into the lattice bulk [Fig.~\ref{fig4} (a), (b)]. 
Under the effect of the index ramp, both states are observed to drift as a whole along the domain wall. The zigzag states exhibit a more profound drift than that in the bearded domain, yet the overall shift for the latter is still measurably observable in term of their COM which we extracted from their intensity distribution. As the plots show, initially, both VHESs move in the positive direction of the $y$-axis and, at a certain distance, arrive at their maximum displacements before moving back. Our results clearly show that the maximum displacement for the light beam in the zigzag domain wall appears much later than in the bearded domain. Comparing their COM finds that when the bearded state completes a full cycle at $z\approx 16~\text{mm}$, the zigzag state has only just reached its maximum displacement. This observation clearly indicates that the period of the BOs for zigzag states is twice that of bearded states, as theoretically predicted. 

This period-doubling effect is further demonstrated in Fig.~\ref{fig4} (c), where we compare the evolution of the COM for both VHESs at three different widths of the induced index ramp $\eta$, i.e., $\eta=120, 90, 60~\mu \text{m}$, 
corresponding to three different gradients $\beta$ of refractive index ~\cite{supplemental}. For each $\eta$, the experiment was conducted five times under the same parameter settings,
thus each curve in Fig.~\ref{fig4}(c)  is the average of the outcomes obtained from five experimental repetitions. 

As expected, a larger gradient of refractive index leads to faster BOs of the wavepackets. Therefore, the period of BOs for zigzag states is about $32~\text{mm}$ at $\eta=120~\mu \text{m}$, but it is reduced to around $20~\text{mm}$ at $\eta=60~\mu \text{m}$. This allows us to observe a complete cycle of BOs within the $2~\text{cm}$ long sample [Fig.~\ref{fig4}(c), right]. However, it is worth noting that the period at $\eta=60~\mu \text{m}$ is not exactly halved compared to $\eta=120~\mu \text{m}$, as would be expected in the ideal case. This discrepancy is attributed to the relatively narrow width of the index ramp ($\eta=60~\mu \text{m}$) compared to the width of the wavepacket, which has a fixed $y$-width of $90~\mu \text{m}$. Due to this, the light beam experiences a slightly shallower ``local" index ramp when oscillating near the ramp boundary, compared to oscillating in the ramp center. As a result, the effective ramp experienced by the light wavepacket is reduced, leading to a slightly longer period. Among others~\cite{supplemental}, the slowly varying gradient of the index also explains the unevenness in the overall oscillation seen in the right two panels of Fig. 4(c), compared to the more ideal oscillation observed in the first panel. Surprisingly, however, although the oscillation speed may slowly change during the oscillation, which could potentially affect the period and amplitude of oscillation, the relative relationship of period doubling between zigzag and bearded domain walls remains fixed. This is specifically illustrated by the vertical dashed blue lines, which represent the distance the light beam at the zigzag domains travels to reach the maximum displacement (half of its oscillation period),  is found to match well with the distance at which the light beam at the bearded domains completes one full oscillation cycle!

Finally, Fig.~\ref{fig4}(c) also reveals a remarkable relationship between the amplitude of BOs for the two VHESs. As their oscillation periods decrease with the decreasing of $\eta$~(or increasing $\beta$ ), their amplitudes also decrease. However, in all the shown cases, the amplitude of oscillation in zigzag states is significantly larger (two times larger) compared to the bearded states, reflecting that the former BOs involve two bands while the latter involves only one.

Before closing, it is worth noting that the glide-reflection symmetry, which originates in the band degeneracy and period-doubling of BOs in this study, is just one aspect of broader nontrivial point-group $G$~\cite{hiller1986crystallography}. Apart from the symmetries of glide of reflection, other symmetries of $G$, including the symmetry of rotations, inversion, and screw rotations, are also often present, that in general indicates BOs with $\upmu$-multiple periods~\cite{hoeller2018topological}. Hence, our experimental observation of period-doubling ($\upmu=2$) marks the first measurement of an entire family of topological invariants, and is expected to promote general measurements of the topological invariants $\upmu$ characterizing the relevant topological materials.
\\
\acknowledgements
N. K. and F. Y. acknowledge the support of Shanghai Jiao Tong University Scientific and Technological Innovation Funds, and Shanghai Outstanding Academic Leaders Plan (No. 20XD1402000). P. W. acknowledges funding from the National Natural Science Foundation (No.~12304366) and China Postdoctoral Science Foundation (No. BX20230217). Q. F. acknowledges the support of China Postdoctoral Science Foundation (No. BX20230218).


\begin{thebibliography}{99}
\bibitem{note} An alternative viewpoint in the interpretation of the band crossing in the zigzag Valley Hall edge states is that the zigzag domain wall itself can be regarded as a standard 1D Su-Schrieffer-Heeger (SSH) model, where the intercell and intracell couplings are equal.
\end{thebibliography}
\end{document}